\documentstyle[preprint,aps,epsf]{revtex}
\global\firstfigfalse
\global\firsttabfalse
\makeatletter
\global\@specialpagefalse
\def\@oddhead{}
\let\@evenhead\@oddhead
\def\@oddfoot{\reset@font\rm \hfill \thepage
{\footnotesize $_{\mbox{(Date:\today)} }$} \hfill}
\let\@evenfoot\@oddfoot
\makeatother

\begin{document}
\draft %\baselineskip=1.5cm
\title{Electron Scattering Through a Quantum Dot: A Phase Lapse Mechanism}
\author{Yuval Oreg and Yuval Gefen}
\address{Department of Condensed Matter Physics\\
The Weizmann Institute of Science, Rehovot 76100 ISRAEL}
\date{\today }
\maketitle

\begin{abstract}
The scattering phase shift of an electron transferred through a quantum dot
is studied within a model Hamiltonian, accounting for both the electron--electron
interaction in the dot and a finite temperature. It is shown that, unlike
in an independent electron picture, this phase may exhibit a phase lapse of $
\pi $ {\em between } consecutive resonances under generic circumstances.
\end{abstract}

\pacs{PACS:73.23.Hk, 73.23.-b, 73.61.-r}

In a recent elegant experiment Yacoby et al. \cite{CB:Yacoby9596}
investigated phase coherence in the Coulomb blockade regime. They have
introduced a new ingredient concerning the physics of a quantum dot,
namely, the scattering phase shift of an electron transmitted through a
dot.  Information about this phase could be obtained by incorporating
the dot in a two probe Aharonov--Bohm interferometer. The experiment
resulted in a remarkable observation: as the gate voltage on the dot
is varied (keeping the Fermi energy of the leads attached to the dot
unchanged), it is possible to scan the phase of the transmission
amplitude over consecutive resonances (in energy). Changing the
location of the Fermi energy from below resonance $1 $ (denote this
situation by $1^- $) to above resonance $1 $ ($1^+$), the phase shift,
 $\theta $, is expected to increase by $\pi $, i.e., $\theta(1^+) -
\theta(1^-) = \pi $. Similarly $\theta(2^+) - \theta(2^-) = \pi $ etc.
The results of the experiment, consistent with these expectations, are
clearly born out by the zero temperature Friedel sum rule \cite
{RFS:Friedel52,RFS:Langer61}. The unexpected part of the observation
was the indication that $\theta(2^-) - \theta(1^+) = \pm \pi $. In
other words, the experiment suggests that there is a phase lapse of
$\pi $ {\em between } two consecutive resonances. This phase lapse has
been observed directly in a recent four probe measurement
\cite{CB:Schuster96}, and has been discussed in a few theoretical
works \cite{CB:Yeyati95,CB:Hackenbroich96,CB:Bruder96s}.

In the present work we propose a mechanism that produces such an
inter--resonance phase lapse. To what extent this is relevant to the
the Coulomb blockade interference experiments
\cite{CB:Yacoby9596,CB:Schuster96,CB:Buks96} is left for future
discussion. We stress, though, that the phase lapse discussed here,
being an inherently finite temperature many body effect, has no
analogue in an independent electron system, and is not born out by the
Friedel sum rule. The phase lapse mechanism discussed here can be
tested in experiments which include tunneling through a two or a few
level system, e.g. a quantum spin.

The first step in our analysis is to construct a simple model for the dot
and the leads attached to it, underlining the essentials required to
observe the phase lapse alluded to above. Consider the following Hamiltonian
(our Hamiltonian is a special case of the one employed in Ref.~\cite{CB:Hackenbroich96})
\begin{equation}
H=H_L+H_{DL}+H_D+H_{DR}+H_R.  \label{modela}
\end{equation}
Here $H_L \: (H_R)$ describes the  leads on the left (right), with
the corresponding Fermi operator ${\hat{L}} \: (\hat{R})$: 
\begin{eqnarray}
H_L &=&\sum_k\varepsilon _L(k)L^{\dagger }(k)L(k)  \label{eq:Hleads} \\
H_R &=&\sum_k\varepsilon _R(k)R^{\dagger }(k)R(k)  \nonumber
\end{eqnarray}
where $k$ runs over the single electron momentum states of the leads. The
dot is modeled as a two--level system 
\begin{equation}
H_D=\varepsilon _aa^{\dagger }a+\varepsilon _bb^{\dagger }b+Ua^{\dagger
}ab^{\dagger }b,  \label{eq:Hdot}
\end{equation}
where $U$ is an interaction term \cite{CB:InterTerm}. Coupling to the leads
is described by

\begin{eqnarray}
H_{DL} &=&V_{aL}a^{\dagger }L(x_1)+V_{bL}b^{\dagger }L(x_1)+h.c.,
\label{eq:Hdot-leads} \\
H_{DR} &=&V_{aR}a^{\dagger }R(x_2)+V_{bR}b^{\dagger }R(x_2)+h.c.,  \nonumber
\end{eqnarray}
where $x_1,x_2$ are two coordinates on the left and on the right leads
respectively, near the contacts to the dots. Here $a^{\dagger },\,b^{\dagger
}$ are the creation operators of the two single electrons states on the dot.
We assume that the coupling to the leads (through tunneling), given by the
quantum hopping terms, is weak. Transfer of an electron through the dot may
be classified into several, qualitatively different, processes 
\cite{CB:Averin90,CB:Aleiner96}. These include sequential tunneling and inelastic
cotunneling, which do not play a role in an interference experiment since a
transfer of an electron through the dot is accompanied by a change in its
quantum state. Here we shall focus on coherent processes, so called elastic
cotunneling.

The coherent transmission {\em amplitude }through the dot is given by a
single electron retarded propagator from the l.h.s lead to the lead on the
r.h.s., $G_{RL}$. The corresponding imaginary time propagator is given by

\begin{equation}
\begin{array}{lll}
{\cal G}_{RL}(\epsilon _n)& =& \displaystyle{ -\int_0^\beta e^{i\epsilon _n\tau }\text{Tr}
\left\{ \exp \left( \beta \left( \Omega -H\right) \right) {\cal T}\left(
R(x_R,\tau )L^{\dagger }(x_L,0)\right) \right\} d\tau} \\
~~ &\equiv& \displaystyle{-\int_0^\beta
e^{i\epsilon _n\tau }\left\langle {\cal T}\left( R(x_R,\tau )L^{\dagger
}(x_L,0)\right) \right\rangle d\tau.}\\
\end{array}
  \label{eq:prop}
\end{equation}
Here $\Omega $ is the Gibbs potential, $\tau =it$ is imaginary time, $
\beta $ is the inverse temperature, $\left\{ \epsilon _n\right\} $ are
fermionic Matsubara frequencies and ${\cal T}$ is the imaginary time ordering
operator. The operators $R$ and $L$ are taken at points $x_R,\,x_L$ on the
r.h.s and l.h.s leads respectively. The retarded propagator $G_{RL}$ is
obtained by performing the analytical continuation  $i\varepsilon _n\rightarrow \varepsilon
+i\gamma ^{+}$ \cite{CB:Gamma}. The propagator ${\cal G}_{RL}$, corresponding
to the coherent transmission amplitude, is now calculated within a
perturbation theory with respect to $H_{DL}$ and $H_{DR}$. To second order
in the coupling it is given by \cite{RFS:AGD63}

\begin{equation}
{\cal G}_{RL}(\varepsilon _n)={\cal G}_R(\varepsilon _n;x_R,x_2)\left[
V_{aR}^{*}{\cal G}_a(\varepsilon _n)V_{aL}{\cal +}V_{bR}^{*}{\cal G}
_b(\varepsilon _n)V_{bL}\right] {\cal G}_L(\varepsilon _n;x_1,x_L),
\label{eq:perprop}
\end{equation}
Here ${\cal G}_L\left( {\cal G}_R\right) $ is the propagator associated with
the l.h.s. (r.h.s) lead when it is uncoupled from the dots; ${\cal G}_a$ is the
propagator of an electron prepared at the state $a$ in the uncoupled dot. One
is allowed to consider low order in perturbation theory since the coupling is
weak and we refer to energies $\varepsilon $ far from the resonances
\cite{CB:SmallParameter}. The factors ${\cal G}_R$, ${\cal G}_L$ in 
Eq.~(\ref{eq:perprop}) are given by $-i\pi \nu e^{ik_F\left( x_R-x_2\right) },-i\pi
\nu e^{-ik_F\left( x_L-x_1\right) }$ where $\nu $ is the density of states
in the leads and $k_F$ is the Fermi momentum. The phase factors $
e^{-ik_Fx_2} $ and $e^{ik_Fx_1}$ cancel out similar phase factors in the
coupling potential amplitudes. This yields

\begin{equation}
{\cal G}_{RL}(\varepsilon _n)=-e^{i\theta _g}\pi ^2\nu ^2\left[ \widetilde{V}
_{aR}^{*}{\cal G}_a(\varepsilon _n)\widetilde{V}_{aL}{\cal +}\widetilde{V}
_{bR}^{*}{\cal G}_b(\varepsilon _n)\widetilde{V}_{bL}\right] ,
\label{eq:perprop1}
\end{equation}
where $\theta _g=k_F(x_R-x_L)$ is a geometrical phase. The
coupling amplitudes $\widetilde{V}$ are equal to the coupling amplitudes $V$ 
divided by the  respective phase factors $ e^{-ik_Fx_2} $ and $e^{ik_Fx_1}.$  
In the absence of a magnetic field $\widetilde V$ can be chosen to be real.
The Green functions associated with the dot
are obtained from definitions similar to Eq.~(\ref{eq:prop}), summing over all
relevant many body states in the dot. We adopt the notation
 $\left|0\right\rangle $ (no electron in the dot), 
$\left| a\right\rangle~\equiv~a^{\dagger}\left| 0\right\rangle,\, 
 \left| b\right\rangle~\equiv~b^{\dagger} \left| 0\right\rangle,\,
 \left|ab\right\rangle~\equiv~a^{\dagger}b^{\dagger }\left| 0\right\rangle $
 for these states with the respective
energies $E_0=0,\,E_a=\epsilon _a-\mu ,\,E_b=\epsilon _b-\mu
,\,E_{ab}=\epsilon _a+\epsilon _b-2\mu +U$. The factor $\mu $ appears in the
expression for the grand canonical energies since populating a state in the dot is
associated with the removal of an electron from the leads whose
electrochemical potential is $\mu $. We now define the probabilities of the
dot to be found (at equilibrium) in either of the states $\left| i\right\rangle =\left|
0\right\rangle ,\,\left| a\right\rangle ,\,\left| b\right\rangle ,\,\left|
ab\right\rangle $,

\begin{equation}
P_i=e^{-E_i/T}/\sum_ie^{-E_i/T}.  \label{eq:probabilities}
\end{equation}
We thus find

\begin{mathletters}
\begin{equation}
{\cal G}_a(\epsilon _n)=(P_0+P_a)\frac 1{i\varepsilon _n-(\epsilon _a-\mu )}
+(P_b+P_{ab})\frac 1{i\varepsilon _n-(\epsilon _a-\mu +U)}  \label{eq:ga}
\end{equation}

\begin{equation}
{\cal G}_b(\epsilon _n)=(P_0+P_b)\frac 1{i\varepsilon _n-(\epsilon _b-\mu )}
+(P_a+P_{ab})\frac 1{i\varepsilon _n-(\epsilon _b-\mu +U)}.  \label{eq:gb}
\end{equation}
\end{mathletters}

To obtain the phase shift we need to measure the interference of the
transmitted amplitude with a reference beam, $t_{ref} $. The latter is
assumed to be energy independent. The magnitude of this interference
term is thus given by $2 \text{Re} \left( t_{ref}^* t_{RL} \right) $ where
\begin{equation}
t_{RL}=-\frac 1{\pi \nu }\int d\varepsilon G_{RL}\left( \varepsilon \right) 
\frac{\partial f\left( \varepsilon \right) }{\partial \varepsilon },
\label{eq:trl}
\end{equation}
and $f$ is the Fermi--Dirac distribution. The retarded propagator $G_{RL}$
is obtained by evaluating the expression in Eq.~(\ref{eq:perprop}) (assuming
that the coupling parameters are practically constant over an energy scale of
the order of the temperature). Analytically continuing to real energies and
performing the energy integration in Eq.~(\ref{eq:trl})  we obtain 
\begin{equation}
\begin{array}{llll}
t_{RL}= & e^{i \theta_g} \nu \displaystyle{ \frac{1}{2iT}} \left\{ {}\right.  & \widetilde{V}_{aR}^{*}\widetilde{V}_{aL}(P_0+P_a)\psi ^{\prime
}((-E_a+i\gamma )/(2\pi iT)+1/2) & + \\ 
&  & \widetilde{V}_{aR}^{*}\widetilde{V}_{aL}(P_a+P_{ab})\psi ^{\prime }((-(E_a+U)+i\gamma )/(2\pi
iT)+1/2) & + \\ 
&  &\widetilde{V}_{bR}^{*}\widetilde{V}_{bL}(P_0+P_b)\psi ^{\prime }((-E_b+i\gamma )/(2\pi iT)+1/2)
& + \\ 
&  & \widetilde{V}_{bR}^{*} \widetilde{V}_{bL}(P_b+P_{ab})\psi ^{\prime }((-(E_b+U)+i\gamma )/(2\pi
iT)+1/2) & \left. {}\right\} ,
\end{array}
\label{eq:tVg}
\end{equation}
where $\psi ^{\prime }\left( z\right) $ is the trigamma function.

Trying to mimic real systems, the states $a $ and $b $ may represent two
 consecutive single particle levels in the dot's spectrum. Their energies
 may depend (linearly) on the gate voltage applied on the dot. Below
 we employ the parameterization
\begin{equation}
  \label{eq:E(Vg)}
  E_a = V_g, \; E_b = V_g + \Delta.
\end{equation}
 
\begin{figure}[htbp]
\leavevmode
\epsfbox{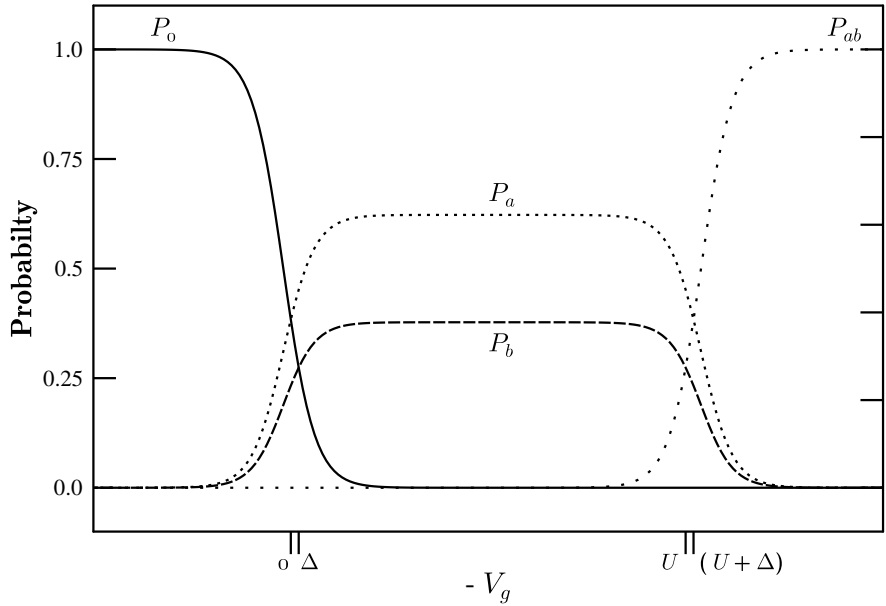}
\caption{{\it  Equilibrium probabilities of the four many--body dot states
                      as a function of $V_g$, $\Delta < T  \ll U\;$ ($\Delta=10\mu eV$,
                     $T=20\mu eV$ and $U=500 \mu eV$.)}}
\label{fg:prob}
\end{figure}

Before we proceed with detailed results, we present a qualitative
discussion pertaining to the essential physics which leads to the 
interresonance phase lapse. Fig.~\ref{fg:prob} presents the various
occupation probabilities as a function of the gate voltage $V_g $, for
 $\Delta < T \ll U$. The range of $V_g$ depicted in Fig.~\ref{fg:prob}
corresponds to two resonances that occur near $V_g=0$ and near $V_g=
- U$, respectively. These resonances, in turn, correspond to a change in 
the state of the dot from  $0 $--occupied to  singly occupied, and from singly
to doubly occupied, respectively.

We note that there are four terms which correspond to four channels
through which an electron transfer through the dot can take place. We
refer to these terms as $A_1$, $A_2 $ (these correspond to the two
terms of ${\cal G}_a $, Eq.~(\ref{eq:ga})) and $B_1 $, $B_2 $.  These
four terms also correspond to the four terms of the transmission
amplitude, Eq.~(\ref{eq:tVg}). We further notice that the terms $A_1$,
 $B_1$ ($A_2$, $B_2\;$) appear in connection with resonance 1
(resonance 2). If $T \ge \Delta $ the thermal factors associated with
 $A_1$ and $B_1$ are of comparable magnitude for the whole range of
 $V_g$ ( a similar statement holds for $A_2$, $B_2$). At the same time,
if our model represents a dot with a random potential (or of a chaotic
shape), the coupling parameters exhibit strong level--to--level
fluctuations \cite{CB:Hackenbroich96,CB:Alhassid95}.
 Let us assume, for example, that for a given sample 
 $\left| V^*_{aR}V_{aL} \right|$ is significantly larger than 
 $ \left| V^*_{bR}V_{bL} \right|$. The main contribution to the transmission
is then due to $A_1$, $A_2$. The term $A_1$ (cf. first term in 
Eq.~(\ref{eq:ga}), and  in Eq.~(\ref{eq:tVg})), represents the sum of two different
second order processes: (1) an electron from the left hops to level
$a$ (which was vacant) and then hops to the right; (2) an electron from
level $a$, (which originally was occupied while level $b$ was empty)
hops to the right, and then an electron from the lead on the left hops to state
$a$. (i.e., a transfer of a hole from right to left.)
 Schematically, these two processes are described by the two
following time sequences of the dot: (1)~$ \left| 0 \right> 
\rightarrow \left| a \right> \rightarrow \left| 0 \right>$, (2)~$
\left| a \right> \rightarrow \left| 0 \right> \rightarrow \left| a
\right>$. In either process the transfer of an electron through the
dot involves the single electron level $a$. 
 A similar analysis applies to the term $A_2$, with two
time sequences (1)~$ \left| b \right> \rightarrow \left| ab \right>
\rightarrow \left| b \right>$, (2)~$ \left| ab \right> \rightarrow
\left| b \right> \rightarrow \left| ab \right>$. Again the electron
transfer takes place through the very same single particle state
$a$. This is unlike an independent electron picture, whereby two
consecutive resonances are associated with two different single
particle states\cite{CB:Inelastic-cotunneling}.

 The coupling factor, $ V^*_{aR} V_{aL} $ in our
example, is common to both $A_1$ and $A_2$. Between the resonances ( and far
from them ) the thermal factors are practically
constant, but the denominators involved in $A_1$ and $A_2$
(Eq.~(\ref{eq:ga})) do vary with $V_g$. The one in $A_1$ (practically
real positive) decreases as $V_g$ becomes more negative, while the
second denominator (practically real negative) increases in magnitude.
At a certain point between the resonances $A_2$ takes over $A_1$,
which introduces a phase lapse of $\pi $.

\begin{figure}[htbp]
\begin{center}
\leavevmode
\epsfbox{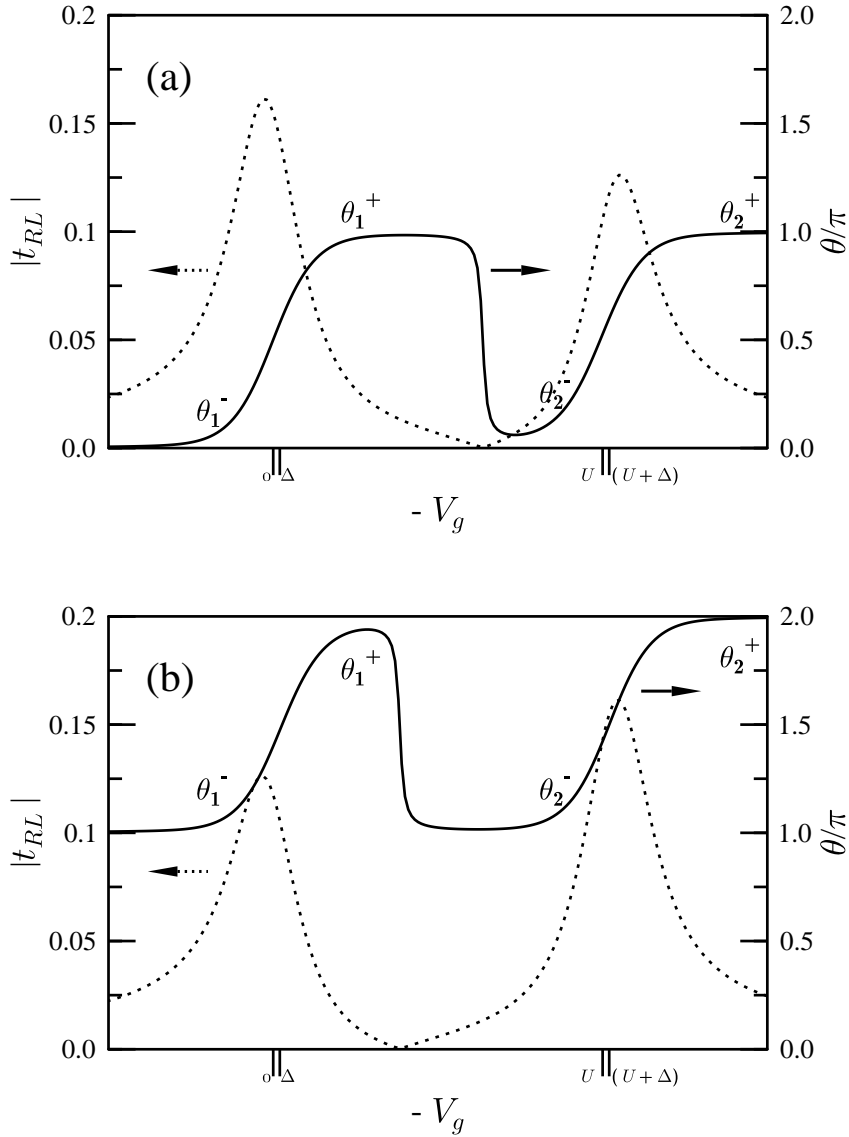}
\caption{{\it The argument of $t_{RL} $ (Eq.~(\protect{\ref{eq:tVg}})) (solid line) 
    and its magnitude (dashed line) for $U=500\mu eV $, $T=20\mu eV$,
    $\Delta = 10 \mu eV$ and $\gamma = 5 \mu eV$ (a) $\widetilde {V}
    _{bL}=-\widetilde{V}_{bR}= 0.3 \widetilde {V}_{aL} =
    0.3\widetilde {V}_{aR}$ (b) $0.3\widetilde {V}_{bL}=-0.3 \widetilde
    {V}_{bR}=  \widetilde {V}_{aL} =  \widetilde {V}_{aR}$.
    The perturbative result (second order the couplings) should not be
    trusted close to the resonances. The interresonance phase lapse is
    apparent.}}
\label{fg:phase-lapse}
\end{center}
\end{figure}

Fig.~\ref{fg:phase-lapse}a depicts the phase and the magnitude of the
transmission amplitude for the case $T > \Delta $, when the coupling to
level $a $ is stronger than the coupling to level $b$. An opposite
case, where the coupling to level $b$ is stronger, is presented in
Fig.~\ref{fg:phase-lapse}b.
For $T > \Delta $ the phase lapse between the resonances is clearly
seen and is robust to changes in the values of the parameters as long as
\begin{equation}
  \label{eq:validity}
\Delta/U  \ll  1 \; \; \text{ and } \; \; 
            \frac{(\widetilde{V}^*_{aR} \widetilde{V}_{aL} - \widetilde{V}^*_{bR} \widetilde{V}_{bL} )}
                   { (\widetilde{V}^*_{aR} \widetilde{V}_{aL} + \widetilde{V}^*_{bR}  \widetilde{V}_{bL} )}   <  \frac{2T}{\Delta}.
\end{equation}

Note that the width of the phase lapse is $ \sim \gamma $ and is hardly affected by 
a finite temperature. The reason is that $G_{RL} (\epsilon) $ is a linear function of
 $\varepsilon $ (in the vicinity of the phase lapse). Performing the energy integration 
in Eq.~(\ref{eq:trl}) results in a virtually temperature independent  $t_{RL} $.
  
\begin{figure}[htbp]
\leavevmode
\epsfbox{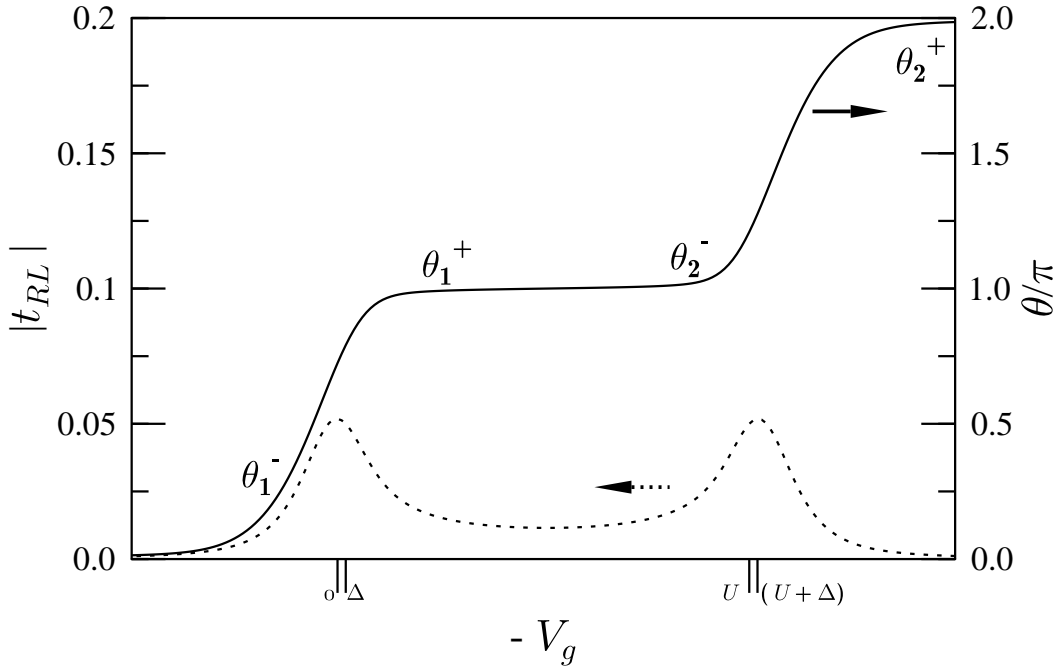}
\caption{{\it The argument of $t_{RL}$ (solid line) and its magnitude (dashed line) for the 
               non generic case $\widetilde{V}_{bL}=-\widetilde{V}_{bR}= 
                 \widetilde{V}_{aL} =  \widetilde{V}_{aR}$, 
              $U=500\mu eV $, $T=10\mu eV$, $\Delta = 10 \mu eV$ and 
               $\gamma = 5 \mu eV$ no phase lapse is observed.}}
\label{fg:nongeneric}
\end{figure}

This phase lapse disappears in the non--generic case where the coupling to level $a$
 and $b$ are exactly equal in magnitude \cite{CB:Sign-of-V}. This is depicted in 
Fig.~\ref{fg:nongeneric} where no interresonance phase lapse is observed \cite{CB:LowTemp}. 

Our results may be generalized to an $n$--level dot. To observe 
an interresonance phase lapse between every two consecutive
 resonances, we require large enough statistical fluctuations of the couplings
 $\{ V \}$ and a finite temperature, which satisfies a generalization of Eq.~(\ref{eq:validity}).
 
Finally, it is interesting to compare our findings with the Friedel sum 
rule \cite{RFS:Friedel52,RFS:Langer61}. The latter, addressing the 
scattering phase shift right at the Fermi energy (hence at zero temperature),
predicts that as we sweep through $n$ resonances, the phase 
shift increases by $n\pi $. Our present analysis addresses a finite temperature scenario,
and therefore is not in contradiction with the Friedel sum rule. (The interresonance phase 
lapse disappears at $T=0$).

In summary, we have considered a few level model Hamiltonian describing
 a dot connected to leads, accounting for a finite temperature and an 
electron--electron interaction.
 The latter leads to a separation of consecutive resonances.
The same single electron level may dominate the electron transfer in two 
resonances that are consecutive in energy. This leads to an interresonance
phase lapse in the scattering phase shift, hence two consecutive resonances
are in--phase vis--a--vis the scattering phase shift. We propose
that this effect may be observed in a transmission experiment through a
few level system (e.g. a quantum spin).

We gratefully acknowledge comments by Y.~Alhassid, A.~M.~Finkel'stein  and A.~Kamenev.
 We have also greatly benefited from discussions with M.~Heiblum,
R.~Schuster, A.~Yacoby and E.~Buks who communicated to us their
results prior to publication.  This research was supported by the
Israel Academy of Sciences, the German--Israel Foundation (GIF) and
the U.S.--Israel Binational Science (BSF).

\bibliographystyle{prsty}
%\bibliography{/home4/fnoreg/motti/manuscript/cb,/home4/fnoreg/ref/library}

\begin{thebibliography}{10}
\bibitem{CB:InterTerm}
$U$ can be thought of as a capacitance term, or, when $a$ and $b$ stand for
  spin states, as a Hubbard term.
\bibitem{CB:SmallParameter}
Far on the scale $V=$ max $ \{ { \nu |V_{aL}|^2}, {\nu |V_{bL}|^2}, {\nu
  |V_{aR}|^2}, {\nu |V_{bR}|^2}, \gamma \}$. To be specific the small parameter
  of the expansion is $V/\delta E$ where $\delta E$ is the distance from the
  resonances. The parameters of the problem will not be renormalized due to
  multiple tunneling processes as long as $(V / \delta E) \log (D / V ) \ll 1$,
  where $D$ is the band width \protect{\cite{KE&AMM:Hewson93}}. We stress that
  the problem addressed here is that of the phase--slip {\em between}
  resonances, not the more trivial effect of a phase--shift by $\pi $ as one
  sweeps through a resonance.
\bibitem{CB:Inelastic-cotunneling}
Note that a sequence such as $ \left| a \right> \rightarrow \left| 0 \right>
  \rightarrow \left| b \right>$ does not contribute to coherent transmission
  since it involves a change of the quantum state of the dot. Such inelastic
  cotunneling processes are not accounted for by the single particle Green's
  function.
\bibitem{CB:Sign-of-V}
We have chosen the relative sign of the parameters $ \widetilde{V}^*_{aR}
  \widetilde{V}_{aL}$ with $ \widetilde{V}^*_{bR} \widetilde{V}_{bL}$ to be
  opposite. As was noted in
  Refs.\protect{\cite{CB:Hackenbroich96,CB:Alhassid95}} this may not always be
  the case, but it does not affect our generic results outlined here.
\bibitem{CB:LowTemp}
Our picture is also modified in the low temperature, $ T < \Delta $ limit,
  where the inter--resonance phase slip may interfere with the phase jump at
  the resonance. This more intricate picture goes beyond the second order
  analysis presented here.
\end{thebibliography}

\end{document}